# Attenuation of the NMR signal due to hydrodynamic Brownian motion

Vladimír Lisý [*], Jana Tóthová

*Department of Physics, Technical University of Košice, Slovakia*

**Abstract.** Nuclear magnetic resonance (NMR) is a widely used nondestructive method to study random motion of spin-bearing particles in different systems. In the long-time limit the theoretical description of the NMR experiments is well developed and allows proper interpretation of measurements of normal and anomalous diffusion. The traditional description becomes, however, insufficient for the shorter-time dynamics of the particles. In the present paper, the all-time attenuation function of the NMR signal in a magnetic-field gradient due to the Brownian motion (BM) of particles in incompressible liquids is calculated by using the method of accumulation of phases by a precessing magnetic moment, without reference to a concrete model of the stochastic dynamics. The obtained expressions are then used to evaluate the attenuation within the hydrodynamic theory of the BM. It is shown that the well-known time behavior of the formulas corresponding to the Einstein theory of diffusion in the case of steady gradient and Hahn's echo experiments is reached at times much larger than the characteristic time of the loss of memory in the particle dynamics. At shorter times the attenuation function significantly differs from the classical formulas used to interpret these experiments.

## 1. Introduction

Nuclear magnetic resonance (NMR) is one of the established methods to study self-diffusion and diffusion of spin-bearing particles in various systems [1-6]. The method uses creating, due to the changes of the particle displacements, a nonuniform frequency of the spin precession in a nonuniform magnetic field, which causes a time-dependent attenuation $S(t)$ of the observed NMR signal. Most often the diffusion coefficients of the particles are extracted from $S(t)$, either after applying a steady field gradient after one rf pulse or by using a technique of rf pulse sequences based on the refocusing principle. Beginning from the works [7, 8], the methods have been also successfully used to describe anomalous diffusion, when the mean square displacement (MSD) of the particles follows the law $Z(t) = Ct^\alpha$ with $\alpha \neq 1$ [4, 9]. The description of the motion of particles in the normal ($\alpha = 1$) or anomalous diffusion regime assumes long times of observation – the times much larger than the relaxation (frictional) time of the particles [10] or the characteristic time of the loss of memory in the particle dynamics. The results of most theoretical studies in the literature correspond to such long times and are inapplicable to shorter times. This is rather surprising in view of a number of experimental investigations that have shown that the chaotic motion of Brownian particles in liquids at shorter times is not consistent with the classical theories by Einstein [11] and Langevin [12]. It has been clearly demonstrated, in the last decade particularly by the method of optical trapping [13-19], that the Langevin equation very well describes the Brownian motion (BM) in fluids if it is modified by replacing the Stokes friction force with the Boussinesq-Basset "history force" [20, 21]. In such a generalized (hydrodynamic) theory of the BM the MSD at $t \to 0$ corresponds to the ballistic motion, $Z(t) \sim t^2$, and at long times, in addition to the Einstein term $2Dt$, where $D$ is the diffusion coefficient, it contains important contributions due to the hydrodynamic memory. So, for a free particle this memory is displayed in the so-called long-time tails that slowly decrease to zero with the increase of time. These peculiarities should be reflected also in the attenuation function of the NMR signal. However, to our knowledge, for the hydrodynamic BM the corresponding calculations are absent. Moreover, there are only a few papers where $S(t)$ has been successfully calculated for all times in other models. In Ref. [22] (see also [2]), the function $S(t)$ was found within the standard Langevin model. This theory, however, has strong limitations: in fact, it is applicable only to Brownian particles with the mass much exceeding that of the surrounding particles (such as the Brownian particles in a gas) and is inappropriate for the BM in liquids [18, 19]. In [22] also the generalized Langevin model with a memory integral (which memory kernel exponentially decreases in time) has been considered, however, its solution was not correct [23]. The same model was recently used to describe the NMR induction signal from spin-bearing particles in an experiment with steady gradient [24]. Both the solution of the generalized Langevin equation (GLE) and the basic expression for the attenuation function proposed in [24] were corrected in [25] (for the subsequent discussion see [26, 27]). In Ref. [28] the standard Langevin model was used to calculate $S(t)$ in the case of steady gradient (called there step gradient) and the attenuation of the echo signal by using the GLE with a memory kernel of the type $\Gamma(t) \sim t^{\varepsilon-1}$ with $0 < \varepsilon < 1$. The latter task was not correctly solved, which can be seen by comparing the long-time limit for $S(t)$ that in [28] does not agree with the known solution for anomalously diffusing particles [8, 23].

In [23] the attenuation of the NMR signal was considered coming from the GLE with the exponentially decreasing memory kernel. This equation is in the literature used to model the Brownian dynamics in viscoelastic (Maxwell) fluids [29, 30]. In the present paper we apply the hydrodynamic model of the BM that has been convincingly proven to well describe the BM in simple incompressible fluids. Before doing this, in the next section we give the basic formulas for the attenuation of the NMR signals. They are expressed through the MSD of the spin-bearing particles in a simple form suitable for calculations without reference to a concrete model of the particle stochastic dynamics, assuming only the Gaussian distribution of the random variables. Then these formulas are used to evaluate the damping of the NMR relaxation signals due to the BM in the case of the steady-gradient and echo experiments.

## 2. NMR signal attenuation due to Brownian motion

Let us first consider an experiment, in which the nuclear induction signal is read-out in the presence of a steady gradient of the magnetic field [24, 31, 32]. A strong magnetic field along the axis $z$ creates the magnetization of spin-bearing Brownian particles in a liquid environment. After the 90° rf pulse the magnetization is modulated by the field gradient of strength $g$ applied at time $t_g$. Within the concept of accumulating phases [33, 7, 8], the deviation of the precessional phase of an individual spin at time $t$ from its phase at time $t_g$ should be calculated as

$$\phi(t) = \int_{t_g}^{t} \omega(t') dt' = -\gamma_n \int_{t_g}^{t} g\left[z(t') - z(t_g)\right] dt', \tag{1}$$

where for the time-dependent resonance frequency offset $\omega(t)$ in the rotating frame the Larmor condition $\omega = -\gamma_n B$ ($\gamma_n$ is the nuclear gyromagnetic ratio) is used that relates the precessional frequency to the applied magnetic field $B = B_0 + gz(t)$, so that $\omega(t)$ is expressed through the difference between the positions $z$ of the spin at times $t$ and $t_g$. The influence of the spin motion on the total magnetization determining the observed NMR signal is expressed by the attenuation function [1, 3, 6, 9, 31-33]

$$S(t) = \langle \exp[i\phi(t)] \rangle. \tag{2}$$

Here, the brackets $\langle ... \rangle$ mean the expectation value of the phase factor as a function of time [34]. Our aim is to express $S(t)$ through the MSD, $Z(t) = \langle [z(t) - z(t_g)]^2 \rangle$. If the distribution of the random variable $\phi(t)$ can be approximated by a Gaussian function, or for a small $\phi$, it follows from Eq. (2) that

$$S(t) = \exp\left[-(1/2)\langle \phi^2(t) \rangle\right]. \tag{3}$$

By using Eq. (1) and assuming stationarity of the random process, in which case the MSD $Z(t)$ is an even function of time, the attenuation can be obtained in the simple, model-independent and suitable for any times form [23, 25],

$$S(t) = \exp\left[-\frac{1}{2}\gamma_n^2 g^2 \int_0^t t' Z(t') dt'\right]. \tag{4}$$

When considering the Hahn pulsed echo experiment [35], the relevant time intervals are $\delta$ (duration of the gradient pulses) and $\Delta$ (the time between the beginnings of the gradient pulses). The result for $t_g + \delta < t < t_e/2$ (up to the second rf pulse at time $t_e/2$, where $t_e$ is the echo time) is the same as Eq. (4) with $t = \delta$. Since the NMR signal is affected by the motion of spins only during the action of the gradients, after the second gradient pulse ($t > \Delta + \delta$) the echo attenuation function $A(\delta, \Delta)$ does not depend on $t$ and is determined only by the parameters $\Delta$ and $\delta$,

$$A(\delta, \Delta) = \left\langle \exp\left\{i\gamma_n g\left[\int_{t_g}^{t_g+\delta} (z(t') - z(t_g)) dt' - \int_{t_g+\Delta}^{t_g+\Delta+\delta} (z(t') - z(t_g)) dt'\right]\right\}\right\rangle. \tag{5}$$

The result of calculations also does not depend on $t_g$,

$$A(\delta, \Delta) = \exp\left\{-\frac{1}{2}\gamma_n^2 g^2 \left[\int_0^\delta dt' \int_0^\delta dt'' Z(t'' - t' + \Delta) - 2\int_0^\delta dt' (\delta - t') Z(t')\right]\right\}. \tag{6}$$

Substituting here $\delta = \Delta = t_e/2$, we obtain the damping of the signal at the echo time $t_e$, in the case of the continuously applied gradient. This is in accordance with the formula obtained from

$$\left\langle \phi^2(t) \right\rangle = \gamma_n^2 g^2 \left\langle \left\{ \int_0^{t_e/2} \left[ z(t') - z(0) \right] dt' - \int_{t_e/2}^{t} \left[ z(t') - z(0) \right] dt' \right\}^2 \right\rangle, \quad (7)$$

for the steady gradient, which is valid at any time $t > t_e/2$,

$$\left\langle \phi^2(t) \right\rangle = \gamma_n^2 g^2 \int_0^t dt'(t' - t_e) Z(t') + 2 \int_0^{t_e/2} dt'(2t' - t) Z(t') + 2 \int_0^{t_e/2} dt' \int_0^{t_e/2} dt'' Z(t' - t''). \quad (8)$$

It is easy to check that Eqs. (4) and (6) for $Z(t) = 2Dt$ give the well-known results in the case of the standard Einstein-Fick theory of diffusion [1], as well as in the case of anomalous diffusion [8, 9], when $Z(t) = Ct^\alpha$ with $\alpha < 1$ corresponding to sub-diffusion and $\alpha > 1$ to super-diffusion [36]. The advantage of the obtained equations is however in the possibility to readily calculate the attenuation of the NMR signal for all times and thus to determine the corrections to the diffusion attenuation functions due to the BM of spin-bearing particles. This was done in [23] for the model of Brownian particles in viscoelastic fluids described by the GLE with an exponentially decaying memory kernel. In what follows we consider another model, in which the BM of particles is described by the hydrodynamic theory.

### 3. Attenuation of NMR signals within the hydrodynamic model of Brownian motion

Within the hydrodynamic theory the BM of particles is described by the GLE, in which the Stokes force is replaced by the Boussinesq force [20], naturally appearing as a solution of the linearized Navier-Stokes equations for incompressible fluids [37],

$$M^* \dot{\upsilon}(t) + \gamma \upsilon(t) + \int_{-\infty}^{t} \Gamma(t-t') \dot{\upsilon}(t') dt' = F_{th}(t). \quad (9)$$

Here, $M^* = M + M_s/2$, $M_s$ is the mass of the fluid displaced by the particle of mass $M$, $\upsilon(t) = \dot{z}(t)$ is the particle velocity along one of the coordinate axes, say $z$, $\gamma$ is the friction coefficient, $\Gamma(t) = \gamma(\tau_R/\pi t)^{1/2}$, and $F_{th}(t)$ is the (colored) thermal noise force [38-41]. The characteristic time $\tau_R = R^2 \rho / \eta$ for a spherical particle of radius $R$ in a liquid of density $\rho$ and viscosity $\eta$ determines how the memory in the particle dynamics is lost in time so that at long times, $t \gg \tau_R$, the particle reaches the (standard) diffusion regime. At shorter times, when $t$ is comparable or smaller than $\tau_R$, the behavior of the particle significantly differs from the diffusive one. The solution of Eq. (9) for the MSD has been obtained long ago [42] (for a review of this and later works see [43]). It can be given the form

$$\frac{Z(t)}{2D} = t - 2\left(\frac{\tau_R t}{\pi}\right)^{1/2} + \tau_R - \tau^* + \frac{1}{\tau^*} \frac{1}{\lambda_2 - \lambda_1} \left[ \frac{\exp(\lambda_2^2 t)}{\lambda_2^3} \operatorname{erfc}(-\lambda_2 \sqrt{t}) - \frac{\exp(\lambda_1^2 t)}{\lambda_1^3} \operatorname{erfc}(-\lambda_1 \sqrt{t}) \right],$$

(10)

where $\tau^* = M^*/\gamma$, $\lambda_{1,2} = -(\tau_R^{1/2}/2\tau^*)[1 \mp (1-4\tau^*/\tau_R)^{1/2}]$, and erfc (.) is a complementary error function [44]. The coefficient $D$ is related to microscopic quantities as $D = (2\rho_p/9\rho)k_B T \tau_R/M = (2\rho_p/\rho + 1)k_B T \tau_R/9M^*$, where $\rho_p$ is the mass density of the particle. The solution (10) has been experimentally verified in a number of works, e.g., [13-19]. It significantly differs from the well-known solution of the standard Langevin equation [12] that is not appropriate for liquids but well corresponds to the BM in gases [18],

$$Z(t) = 2D\{t - \tau[1 - \exp(-t/\tau)]\}, \quad \tau = M/\gamma. \tag{11}$$

The attenuation function (4) of the NMR induction signal due to the BM can be obtained exactly by integrating Eq. (10). By using the series representation of the error function [44] one finds $\int_0^t \exp(z^2 t')\text{erfc}(-z\sqrt{t'})dt' = z^{-2}[\exp(z^2 t)\text{erfc}(-z\sqrt{t}) - 1] - 2t^{1/2}/\sqrt{\pi}z$. With the help of this result the integral $\int_0^t t'Z(t')dt'$ can be calculated by parts. Its long-time approximation for $t \gg \tau^*, \tau_R$ gives

$$S(t) \approx \exp\left\{-\frac{\gamma_n^2 g^2 D t^3}{3}\left[1 - \frac{12}{5}\left(\frac{\tau_R}{\pi t}\right)^{1/2} + \frac{3}{2}\frac{\tau_R - \tau^*}{t} + \left(\frac{9}{\pi}\right)^{1/2}\frac{(\tau_R - \tau^*)(\tau_R - 3\tau^*)}{\tau_R^2}\left(\frac{\tau_R}{t}\right)^{5/2}\right]\right\}. \tag{12}$$

This expression is more easily obtained from the long-time expansion of Eq. (10), if only the terms growing with $t$ are taken into account. A comparison with the result for the standard Langevin model (11),

$$S(t) = \exp\left\{-\gamma_n^2 g^2 D t^3\left[\frac{1}{3} - \frac{\tau}{2t} + \left(\frac{\tau}{t}\right)^3 - \left(\frac{\tau}{t}\right)^2\left(1 + \frac{\tau}{t}\right)\exp\left(-\frac{t}{\tau}\right)\right]\right\}, \tag{13}$$

shows that within the hydrodynamic model the classical result based on the Einstein theory of diffusion is reached much more slowly.

At very short times the motion is ballistic, $X(t) \approx k_B T t^2/m$, and Eq. (4) gives

$$S(t) \approx \exp\left[-k_B T \gamma_n^2 g^2 t^4/8m\right], \tag{14}$$

where $m = M^*$ or $m = M$, depending on which of the two models is considered.

In the case of Hahn's echo, Eqs. (10) and (6) give for the attenuation function at long times

$$-\frac{\ln A(\delta, \Delta)}{\gamma_n^2 g^2 D} \approx \delta^2\left(\Delta - \frac{\delta}{3}\right) - \frac{8}{15}\left(\frac{\tau_R}{\pi}\right)^{1/2}\left[(\Delta + \delta)^{\frac{5}{2}} - 2\Delta^{\frac{5}{2}} - 2\delta^{\frac{5}{2}} + (\Delta - \delta)^{\frac{5}{2}}\right]. \tag{15}$$

For the steady gradient we have to take $\Delta = \delta$ and obtain a correction to the Stejskal-Tanner result [45] at the echo time $2\delta$

$$A(\delta, \Delta \approx \delta) \approx \exp\left\{-\frac{2}{3}\gamma_n^2 g^2 D \delta^3 \left[1 - \frac{16}{5}(2^{1/2} - 1)\left(\frac{\tau_R}{\pi\delta}\right)^{1/2}\right]\right\}. \tag{16}$$

When $\delta \to 0$, the echo attenuation from (15) can be approximated by the formula $A(\delta, \Delta) \approx \exp[-\gamma_n^2 g^2 \delta^2 Z(\Delta)/2]$. Here (cf. Eq. (10), Eq. (15) at $\delta \ll \Delta$ determines an important correction to the result known for normal diffusion,

$$A(\delta \ll \Delta, \Delta) \approx \exp\left\{-\gamma_n^2 g^2 D \delta^2 \Delta \left[1 - 2(\tau_R/\pi\Delta)^{1/2}\right]\right\}, \tag{17}$$

and shows a difference from the attenuation within the standard Langevin theory of the BM, which at long times ($\delta \gg M/\gamma$) is [22, 28, 23]

$$-\ln A(\delta, \Delta) \approx (\gamma_n g)^2 D\left[\delta^2(\Delta - \delta/3) - 2\tau^2 \delta\right]. \tag{18}$$

The obtained analytical results are illustrated by Figs. 1 – 3 that present numerical calculations [46] of the MSDs in the considered models and the corresponding attenuation functions. Figure 1 demonstrates the difference between $Z(t)$ given by Eqs. (10) and (11), and the Einstein result $Z(t) = 2Dt$.

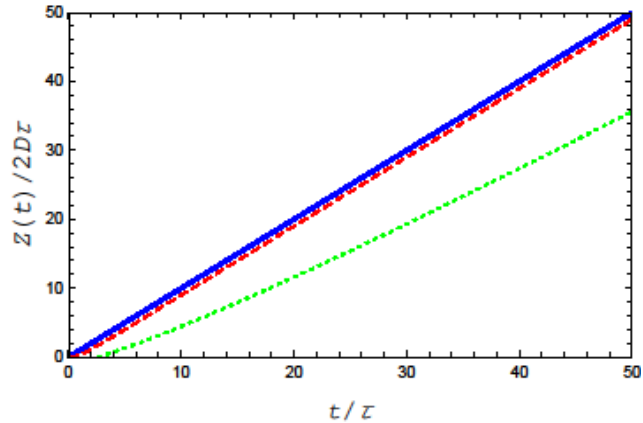

**Fig. 1.** Demonstration of the differences between the exact result for the MSD within the hydrodynamic theory of the Brownian motion, Eq. (10) (dotted line, green online), and the results from the standard Langevin (Eq. (11), dashed line, red online) and Einstein (solid line, blue online) theories. It is assumed that the density $\rho$ of the liquid is equal to that of the particle ($\rho_p$). The MSDs are normalized to $2D\tau$, where $\tau = M/\gamma$ is taken as the unit of time.

Figure 2 illustrates how the classical result from the Einstein diffusion theory, $S_E(t) = \exp(-\gamma_n^2 g^2 D t^3/3)$, is approached with increasing $t$ within the standard Langevin theory and in the theory of the hydrodynamic BM. The results from these theories, Eqs. (12) and (13), are plotted as $\ln S(t)/\ln S_E(t) = -3\ln S(t)/[(\gamma_n g)^2 D t^3]$, so that the classical result corresponds to 1. The time is expressed in dimensionless quantity $t/\tau$. For simplicity, in all figures it is assumed that the density $\rho$ of the liquid is equal to that of the particle ($\rho_p$), as for buoyant Brownian particles. In this case $\tau = 2\tau_R/9$ (it was used that $\tau = M/\gamma$, $\gamma = 6\pi\eta R$, and $M = 4\pi R^3 \rho_p /3$), and $\tau^* = M^*/\gamma = \tau_R/3 = 3\tau/2$. It is seen that the classical result is in the hydrodynamic model reached more slowly (basically as $\sim (\tau/t)^{1/2}$) than in the Langevin model (as $\sim \tau/t$).

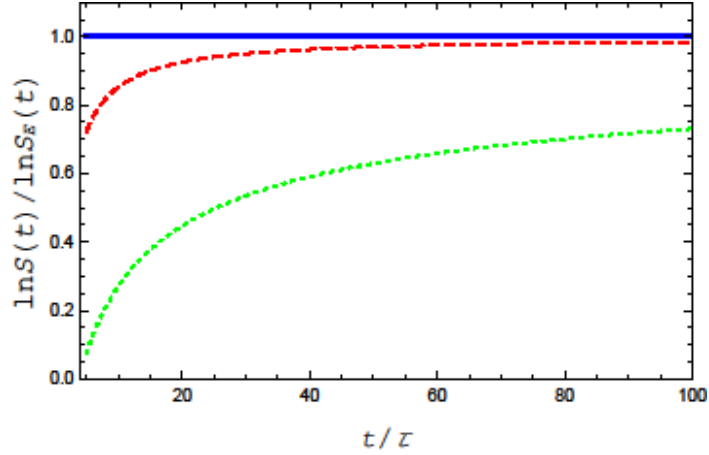

**Fig. 2.** Comparison between the time dependence of the attenuation functions due to the Brownian motion in the steady-gradient experiment described within the Einstein theory (solid line, blue online), standard Langevin theory (Eq. (13), dashed line, red online), and the hydrodynamic model (Eq. (12), dotted line, green online). The numerical calculations correspond to the long-time behavior of $S(t)$ as described in the text, with normalization to the result based on the Einstein theory, which is thus presented as unity.

Even a more significant difference between the two models is demonstrated in Fig. 3 for the echo signal, when the classical result is $A(\delta,\Delta) = \exp[-\gamma_n^2 g^2 D \delta^2 (\Delta - \delta/3)]$. We assume that $\Delta \approx \delta$ and compare the Langevin model, Eq. (18), with Eq. (15). It can be concluded that if the diffusion coefficient is determined from experimental data, the error in its obtaining is time-dependent and can be large. For example, if $D$ would be measured in the experiment with steady gradient, even for $t/\tau$ as large as $10^3$ the difference between its value extracted by using the classical formula and the one from a more realistic hydrodynamic theory is predicted to be still about 10%.

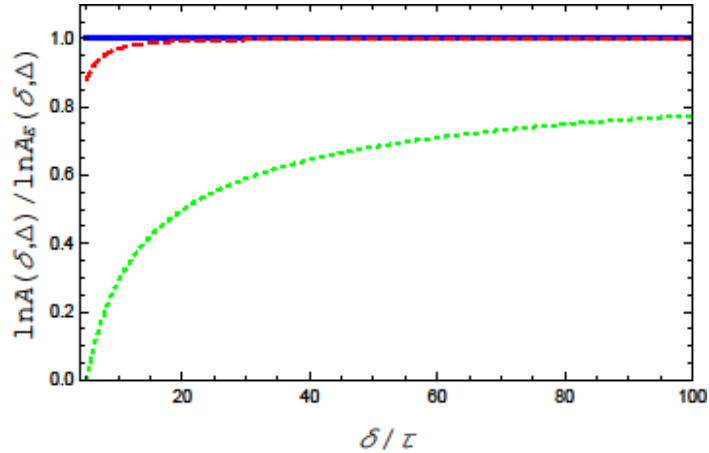

**Fig. 3.** Comparison between the time dependence of the attenuation functions due to the Brownian motion in the steady-gradient spin-echo experiment described within the Einstein theory (solid line, blue online), the hydrodynamic model (Eq. (16), dotted line, green online), and the standard Langevin theory (Eq. (18), dashed line, red online). The numerical calculations correspond to long times as described in the text, with normalization to the result based on the Einstein theory, which is thus presented as unity.

## 4. Conclusion

Nuclear magnetic resonance (NMR), being nondestructive and noninvasive, is a widely used method to study diffusion of spin-bearing particles in different systems, including the biological ones. In the long-time limit, when the particles are in the diffusion regime, the theoretical description of the NMR experiments is well developed and allows to properly interpret measurements of normal and anomalous diffusion. The traditional description becomes, however, insufficient if the interest is in shorter-time dynamics of the particles. A theory is needed that would satisfactory describe the stochastic motion not only at long times. The standard Langevin theory gives the description of the Brownian motion at long times but it does not take into account the memory in the particle dynamics and its applicability is thus very limited: for liquids this theory is improper. It has been shown in a number of works that the Langevin theory for the Brownian motion in simple liquids should be replaced by the hydrodynamic theory, in which the Langevin equation in addition to the Stokes force contains a convolution of the acceleration of the particle with a memory kernel that depends on time as $\sim t^{-1/2}$. In the present paper, the attenuation function of the NMR signal due to the Brownian motion has been evaluated. We considered the experiments when the nuclear induction signal is measured in the presence of a steady field gradient, and the Hahn spin echo (the stimulated echo is obviously described by the same formulas depending only on the duration of the gradient pulses and the time interval between the ends of the gradient pulses). The attenuation is calculated through the accumulation of the spin phases in the frame rotating with the resonance frequency. Coming from the changes of the phases during the time of observation, this accumulation is represented through the mean square displacement for stationary and Gaussian random processes. The obtained formulas give known results in the case of normal and anomalous diffusion, but are equally applicable for systems described by other models, e.g., by the standard Langevin equation or its popular generalizations with the dissipative force represented by a convolution of the particle velocity (or acceleration) and a memory kernel. The expressions for the attenuation function that we propose for the hydrodynamic model significantly differ from the known results based on the standard Langevin theory. The long-time limit within the Einstein diffusion theory is approached very slowly. Due to this our corrections to the classical results should be observable. This could have an impact on the determination of the diffusion coefficient of particles, even if they are extracted in experiments with durations of the gradient pulses (typically milliseconds) much larger than the characteristic time of the loss of memory of the Brownian particle (e.g., $\tau_R$ is of order of $10^{-6}$ s for micrometer-sized particles in water).

## Acknowledgement

This work was supported by the Agency for the Structural Funds of the EU within the project NFP 26220120033, and by the Ministry of Education and Science of the Slovak Republic through grant VEGA No. 1/0348/15.